\documentclass[usenatbib,conference]{basi}

\usepackage[british]{babel}
\usepackage[varg]{txfonts}
\usepackage{rotating}
\usepackage{geometry}                % See geometry.pdf to learn the layout options. There are lots.
\usepackage{graphicx}
\usepackage{amssymb}
\usepackage{epstopdf}
\usepackage{dcolumn}

\begin{document}
\title[Sunspot Seismology]{Sunspot seismology: accounting for magnetohydrodynamic wave processes using imaging spectropolarimetry}
\author[S.~P.~Rajaguru]%
       {S.~P.~Rajaguru\\
       Indian Institute of Astrophysics, Bangalore--560034, India}
\pubyear{2011}

\volume{00}

\date{Received \today}

\maketitle
%------------------------------------------------------------------------------%
% abstract and keywords                                                        %
%------------------------------------------------------------------------------%

\label{firstpage}

\begin{abstract}
The effects of acoustic wave absorption, mode conversion and transmission by a sunspot on the helioseismic
inferences are widely discussed, but yet accounting for them has proved difficult for lack of a consistent
framework within helioseismic modelling. Here, following a discussion of problems and issues that the 
near-surface magnetohydrodynamics hosts through a complex interplay of radiative transfer, measurement issues,
and MHD wave processes, I present some possibilities entirely from observational 
analyses based on imaging spectropolarimetry. In particular, I present some results on wave evolution as
a function of observation height and inclination of magnetic field to the vertical, derived from a high-cadence
imaging spectropolarimetric observation of a sunspot and its surroundings using the instrument IBIS (NSO/Sac Peak,
USA). These observations were made in magnetically sensitive (Fe I 6173 A) and insensitive (Fe I 7090 A) upper 
photospheric absorption lines. Wave travel time contributions from within the photospheric layers
of a sunspot estimated here would then need to be removed from the inversion modelling procedure, that does
not have the provision to account for them. 
\end{abstract}


%------------------------------------------------------------------------------%

% main text of the paper, using \section, \subsection, \subsubsection          %

%------------------------------------------------------------------------------%

\section{Introduction}

Developments in sunspot seismology trace back to the original suggestion
by \citet{1982Natur.297..485T} that interactions between sunspots and helioseismic
p modes could be used to probe the sub-surface structure of sunspots.
The analyses that followed \citet{1982Natur.297..485T} focussed mainly on
changes in the frequency - wavenumber spectrum ($\nu $ - k) and in the
modal power distribution. These studies led to the
discovery of 'absorption' of p modes by sunspots \citep{1987ApJ...319L..27B}:
about 50\% of the flux of acoustic wave energy impinging on a
sunspot is not observed to return to the quiet Sun.
%The subsequent efforts at modelling the magnetic field - p mode
%interactions focussed on understanding and explaining the mechanisms
%responsible for the absorption of p modes by sunspots \citep{1992ApJ...391L.109S,
%1996BASI...24..211S,2003MNRAS.346..381C,2005SoPh..227....1C}.
Development of several local helioseismic techniques, viz.
the ring diagram analysis \citep{1988ApJ...333..996H}, helioseismic holography \citep{1990SoPh..126..101L}
and time-distance helioseismology \citep{1993Natur.362..430D}, has since
brought in new ways of probing the subsurface structure and dynamics of
sunspots. However, the question, viz. is a sunspot formed, in the sub-surface layers, 
of a monolithic flux tube or a cluster of flux tubes?, still remains to be answered.
An answer to this question would also address the dynamics of heat and material flow
in and around sunspots, and hence would have far reaching implications for the
magnetohydrodynamics of solar and stellar magnetism.

Applications of time-distance helioseismology 
appeared as a promising avenue with its 3-dimensional tomographic images of flow
and sound speed structures beneath sunspots \citep{2000SoPh..192..159K,2001ApJ...557..384Z}.
These early results showed an increased
sound speed region extending from about 4 Mm down to about 18 Mm with a maximum
change of about 1 - 2 \%, while the near surface layers in the 1 - 3 Mm depth range
show a decrease in sound speed of similar magnitude. The flow pattern \citep{2001ApJ...557..384Z}
consists of a shallow (1.5 - 3.0 Mm) converging flow that feeds a strong downflow
beneath the sunspot \citep{1996Natur.379..235D} up to depths of about 5 Mm.
Though these results have features indicative of the cluster model \citep{1979ApJ...230..905P},
new developments and improvements in several different fronts in local helioseismology
have served to emphasise the inadequacy of such analyses \citep{gizonetal10}. 
%Especially, inversions of f mode travel 
%times \citep{2000JApA...21..339G} have shown, in agreement with surface measurements from local correlation tracking,
%the presence of an outward moat flow from the sunspots, and 
In contrast to results from time-distance helioseismology, studies based on
phase sensitive holography \citep{2000SoPh..192..307B} have shown phase shifts of waves consistent
with a faster propagation in the near surface layers, in direct correlation with the
surface magnetic proxies (e.g. LOS magnetogram signals), and which decrease monotonically
with depth becoming undetectable at layers deeper than about 5 Mm.
Recent new ways of travel time measurements
and inversions \citep{2010SoPh..267....1M,2011A&A...530A.148S} show that the moat outflows around sunspots extend much
deeper (up to about 4 - 5 Mm).
These new developments have brought to
the fore the dominant direct interactions between acoustic waves and magnetic fields, which
leave too large a signal in measurements to be treated with the conventional
methods of seismic inversions that club such effects into thermal perurbations.

The early contentions that p mode absorption of sunspots could be used to probe them, thus,
have come around a full circle to the realization, through theoretical attempts at
explaining the above surface effects \citep{2003MNRAS.346..381C,2005SoPh..227....1C,
2006ASPC..354..244S}, that they are the very processes that need to be accounted for
before we proceed further in the application of the later developed local helioseismic techniques.

\section{New developments: unreliability of old results}                   \label{spr-sec:issues}

The physical setting and nature of changes in the global structure and dynamics
of the Sun is consistent with a conventional helioseismic analysis procedure, viz. a linear
first order perturbation to the equilibrium structure of the Sun that the p mode frequencies
effectively sense. However, such a treatment is much less adequate to probe the influence of
sunspots in the near surface layers.
The major sources of inadequacy in the local helioseismic analyses of sunspots, as have been
gleaned from recent research, can be identified to arise from
two basic causes: (i) inadequate understanding and modelling of the
interactions between the acoustic waves and the sunspots, where magnetohydrodynamic effects
dominate, and (ii) inadequate identification of the helioseismic observables due to
complexities in the observation and measurement procedures themselves.
However, much of what are known as 'surface magnetic effects' contain subtle inter-mixture
of physical, measurement and analysis issues and hence have contributions from both the
above causes.

\subsection{Surface magnetic effects}                   \label{spr-sec:issues-i}
The neglect of direct magnetic effects
due to the pressure and tension forces of the magnetic field on the wave speed,
while inverting either the travel times or frequency shifts, is the foremost of issues
arising from the cause (i) above. 
Interesting, but not yet fully understood, revelations on the seismic disguises of
the dominantly near-surface interactions between the magnetic field and acoustic
waves came forth from analyses based on phase-sensitive holography \citep{2000SoPh..192..307B}:
the 'showerglass effect' of \citet{2005ApJ...620.1107L} and the 'inclined magnetic field effect'
of \citet{2005ApJ...621L.149S}. The former effect refers to strong surface phase perturbations
that the upcoming acoustic waves in active regions undergo resulting in impairment of their
coherence, similar to the blurring of images seen through a commercial showerglass.
These are measured as phases of the so called 'local control correlations'
of $ingressing$ (ingoing) and $egressing$ (outgoing) waves with wavefield observed
at a particular point in active region (or sunspot) and are found to increase almost
exponentially with magnetic field strength $B$. The 'inclined magnetic field effect' pertains to the
penumbral regions, where there are anomalous changes that depend on the inclination angle
of the magnetic field and the line of sight angle of observations \citep{2005ApJ...621L.149S}.
To correct for the showerglass effect \citet{2005ApJ...620.1118L} devised a magnetic
proxy, which is a complex amplitude that depends on $B^{2}$ and is nothing but the
reciprocal of the appropriate local control correlation. 'Corrected measurements' follow
upon multiplying local surface signal with the above proxy. Such corrections
\citep{2005ApJ...620.1118L} show that the sub-surface acoustic anomalies disappear
below a depth of $\approx 5$ Mm, in contrast to the time-distance helioseismic inferences
\citep{1996Natur.379..235D,2000SoPh..192..159K}.

It is also likely that additional effects
such as changes in the path length of the waves due to thermal expansion or contraction
make significant contributions of either sign, and hence incorrect estimates of
changes in sound speed from those in travel times. In particular, the path length changes
associated with the Wilson depression and the propagating nature of
(magneto-)acoustic waves (due to reduced cut-off frequency) would add contributions of
opposite signs in the wave travel times. Clearly, neither thermal nor magnetic perturbations
alone can explain the inferences.
 
\subsection{Observational issues: radiative transfer effects}             \label{spr-sec:issues-ii}

The altered thermal conditions in sunspots mean that the transfer of spectral line radiation is
different from that in quiet Sun, and in the case of Zeeman sensitive lines the polarization and shape 
of the line interfere with the Doppler measurement procedure.
With the added situation that the character of waves also
are changed due to the magnetic field, radiative
transfer effects manifest in Doppler velocity signals through subtle
interaction of the above changes: the second basic cause [case (ii)] described earlier, viz.
inadequate identification of the helioseismic observables due to
complexities in the observation and measurement procedures themselves.
Here the helioseismic observable is the phase or travel time of a wave observed within a sunspot.
Because the magnetic field lowers the acoustic cut-off frequency, and because it converts
some of the incident acoustic waves into upward propagating ones confined to follow the field lines,
the phases of waves measured within sunspots depend sensitively on the height within the line forming
layers. Any line of sight angle dependent changes in (line) optical depth would then manifest as
changes in the wave phases, i.e. different locations within a sunspot located at on off-disk center
position would yield different phases for waves. This radiative transfer effect has been
brought out clearly in an observational study of a sunspot in Ni I line (6768 A) using the Advanced
Stokes Polarimeter (ASP) at the Dunn Solar Telescope of the National Solar Observatory at Sac Peak,
Sunspot, New Mexico \citep{rajaguruetal07}.

\section{Helioseismic signatures of wave evolution in the observable layers: an example} \label{spr-sec:nresult}

As an example of the near-surface effects discussed in the previous Section, I present here results of
a study using a high cadence imaging spectropolarimetric observation of a sunspot and its surroundings in magnetically
sensitive (Fe {\sc i} 6173 \AA~) and insensitive (Fe {\sc i} 7090 \AA~) upper photospheric absorption lines.
The results of this study have already been published \citep{rajaguruetal10} and we refer the readers to this
original paper for a detailed account. We restrict ourselves here to a brief account of the observations and
major physical implications of the results. 
The observations were made using the Interferometric BI-dimensional Spectrometer (IBIS)
installed at the Dunn Solar Telescope of the National Solar Observatory, Sac Peak, New Mexico, USA. 
% IBIS has spectral and
%spatial resolutions of 25 m\AA~ and 0".165, respectively, and has a 80" diameter ($\approx$60 Mm) circular field of view (FOV).
We observed a medium sized sunspot (NOAA AR10960, diameter
$\approx$ 18 Mm) located close to the disk center (S07W17) on 2007 June 8. Our observations involved
scanning and imaging in all the Stokes profiles ($I,Q,U,V$) of magnetic Fe {\sc i} 6173.34 \AA~ and in Stokes $I$ of non-magnetic
Fe {\sc i} 7090.4 \AA~, with a cadence of 47.5 s. A 7 hr continuous observation was chosen for our analysis

Line-of-sight (LOS) velocities of plasma motions
within the line forming layers are derived from the Doppler shifts of line bisectors.
We use 10 bisector levels with equal spacing in line intensity, ordered from the line core (level 0)
to the wings (level 9), and derive 10 velocity data cubes, $v_{i}(x,y,t)(i=0,...,9)$, for each line.
For the magnetic line, we use the average of bisector
velocities from the left ($I+V$) and right ($I-V$) circular polarization (CP) profiles
\citep{sankar-rimmele02,dtiniesta03} and those from the $I$ profile for the non-magnetic line.
The 10 bisector levels span the height range within the line formation region in an unique
one-to-one way.

\subsection{Instantaneous Wave Phases and Helioseismic Travel Times}

Instantaneous wave phases in the form of phase shifts
$\delta\phi_{i,0}(\nu)$=Phase[${\mathbf V_{i}}(\nu){\mathbf V^{*}_{0}}(\nu)$], where $\nu$ is the cyclic frequency of a wave and
${\mathbf V}$ is the Fourier transform of $v$, due to wave progression between two heights
corresponding to any one of the bisector levels $i=1,2,...,9$ and level $0$ (the top most layer) are
calculated \citep{rajaguruetal07}. 
The 10 different data cubes from each line are run through a standard $p$-mode time-distance analysis procedure
in center-annulus geometry \citep{rajaguruetal04}. Here, travel times for $\Delta = 16.95$ Mm are analysed,
because, given the sizes of observed region (radius $\approx$ 29 Mm) and the spot (radius $\approx$ 9 Mm), this is the 
optimum $\Delta$ that facilitates distinguishing clearly the ingoing and outgoing waves in the sense of their 
interactions with the spot. Height dependent contributions to out- and ingoing
phase travel times $\tau^{+}$ and $\tau^{-}$ from within the line forming layers are determined using
$\delta\tau^{\pm}_{i,0}=\tau^{\pm}_{0}-\tau^{\pm}_{i}$ ($i=1,...,9$).

\begin{figure}[ht]
\centering
%\figurenum{2}
\centerline{\includegraphics[width=7cm]{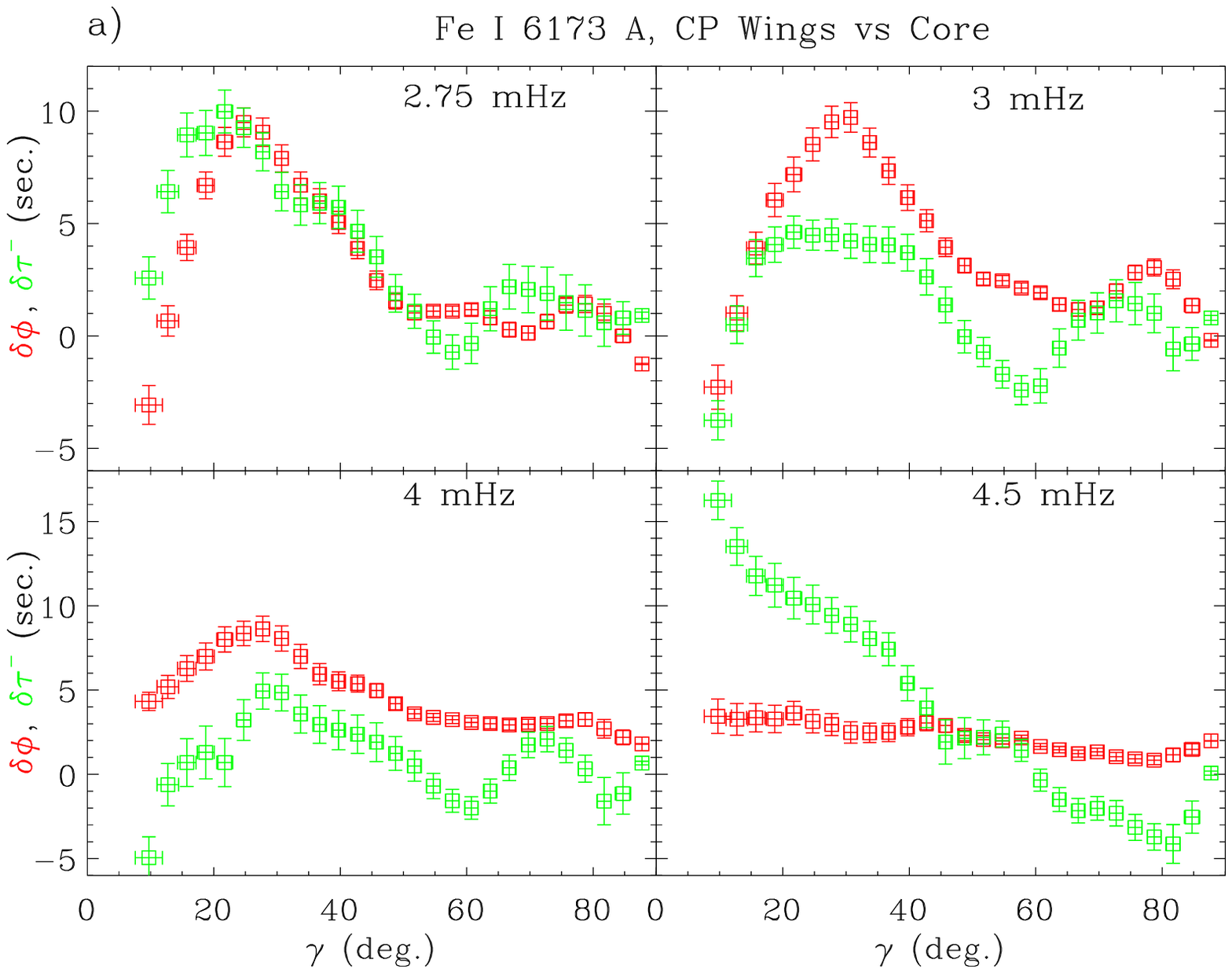} \qquad
            \includegraphics[width=7cm]{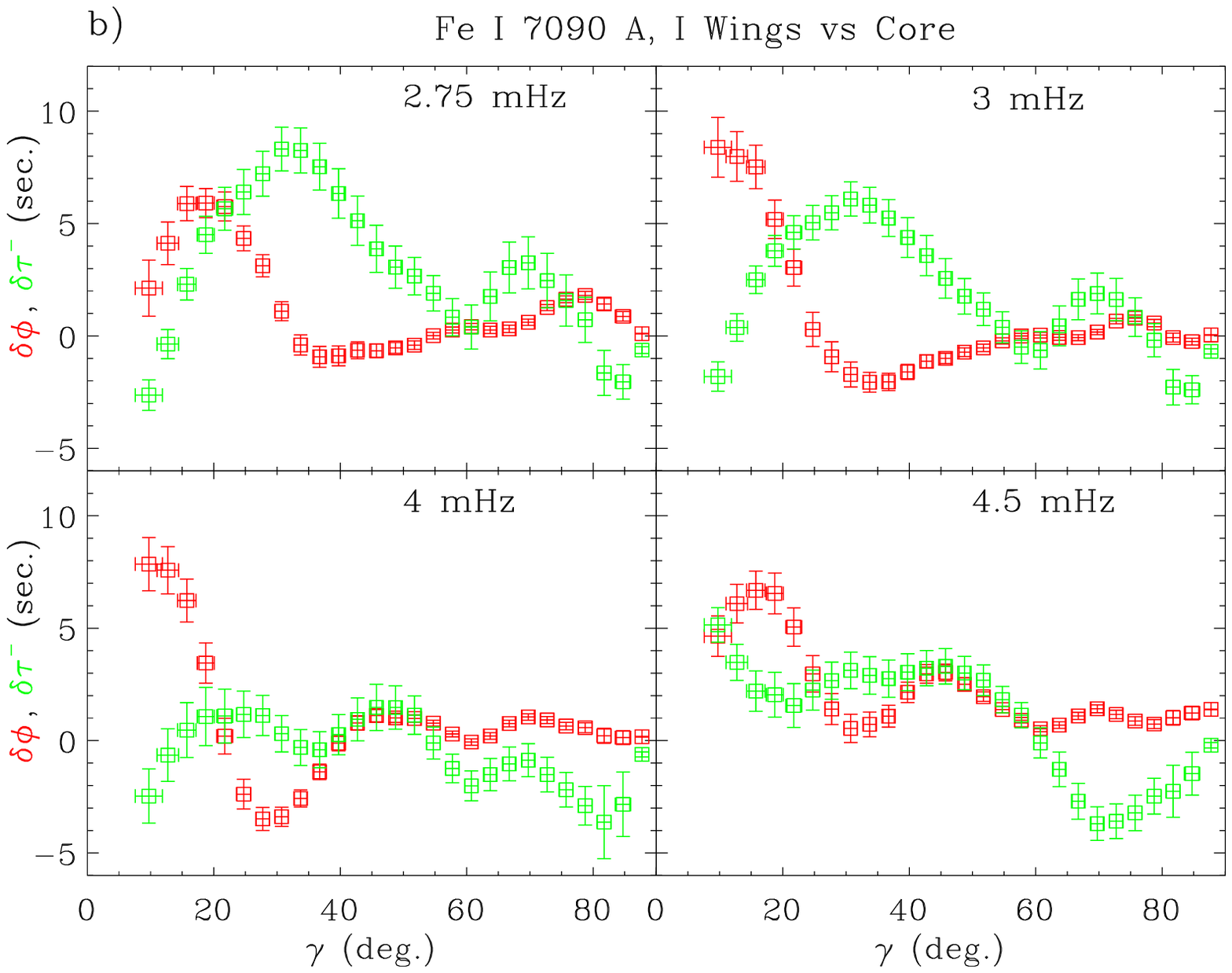}}
%\plottwo{f2a.eps}{f2b.eps}
\caption{Instantaneous phase shifts, $\delta\phi_{8,0}(\nu)$, and changes in ingoing wave travel times,
$\delta\tau^{-}_{8,0}$, due to wave propagation between the formation heights of wings (20 km) and core (270 km)
of Fe {\sc i} 6173 \AA~({\em panel a}), and of Fe {\sc i} 7090 \AA~({\em panel b}) against $\gamma$ of B.}
\label{fig:2}
\end{figure}
We show in Figure 1 $\delta\phi_{8,0}$ and $\delta\tau^{-}_{8,0}$, due to wave evolution within the
region bounded by the wing (level 8) and core (level 0) formation heights, against $\gamma$.
The $\nu$ values marked in the panels of Figure 1 are the central
frequencies of 1 mHz band filters used. Keeping in mind that $\delta\phi_{8,0}$ have contributions from a larger
set of waves (as discussed above), results in Figure 1(a) for the magnetic line show a surprising
amount of correlation between the two measurements, and moreover exhibit a strikingly similar $\gamma$ dependence.
These results immediately reveal several interesting aspects of magnetic field - acoustic wave interactions:
(1) first of all they confirm that helioseismic waves incident on the sunspot see themselves through to higher
layers of its atmosphere with a striking dependence on $\gamma$: a coherent let
through of incident waves happen, peaking around $\gamma \approx$ 30$^{\circ}$, maintaining
a smooth evolution of time-distance correlations; (2) remembering that CP profiles of the magnetic line have maximum
sensitivities for velocities within vertical magnetic field, it is seen that a large fraction of waves propagating upward
within such field are due to helioseismic waves originating at distant locations;
%and hence account substantially for the wave energy absorbed by the sunspot;
and, (3) provide direct evidences that ingoing wave travel times would cause observing height
dependent signals in flow inferences from travel time differences.

\begin{figure}[ht]
%\epsscale{1.1}
\centering
%\figurenum{3}
%\plotone{f3a.eps}
%\plottwo{f3a.eps}{f3b.eps}
\centerline{\includegraphics[width=7cm]{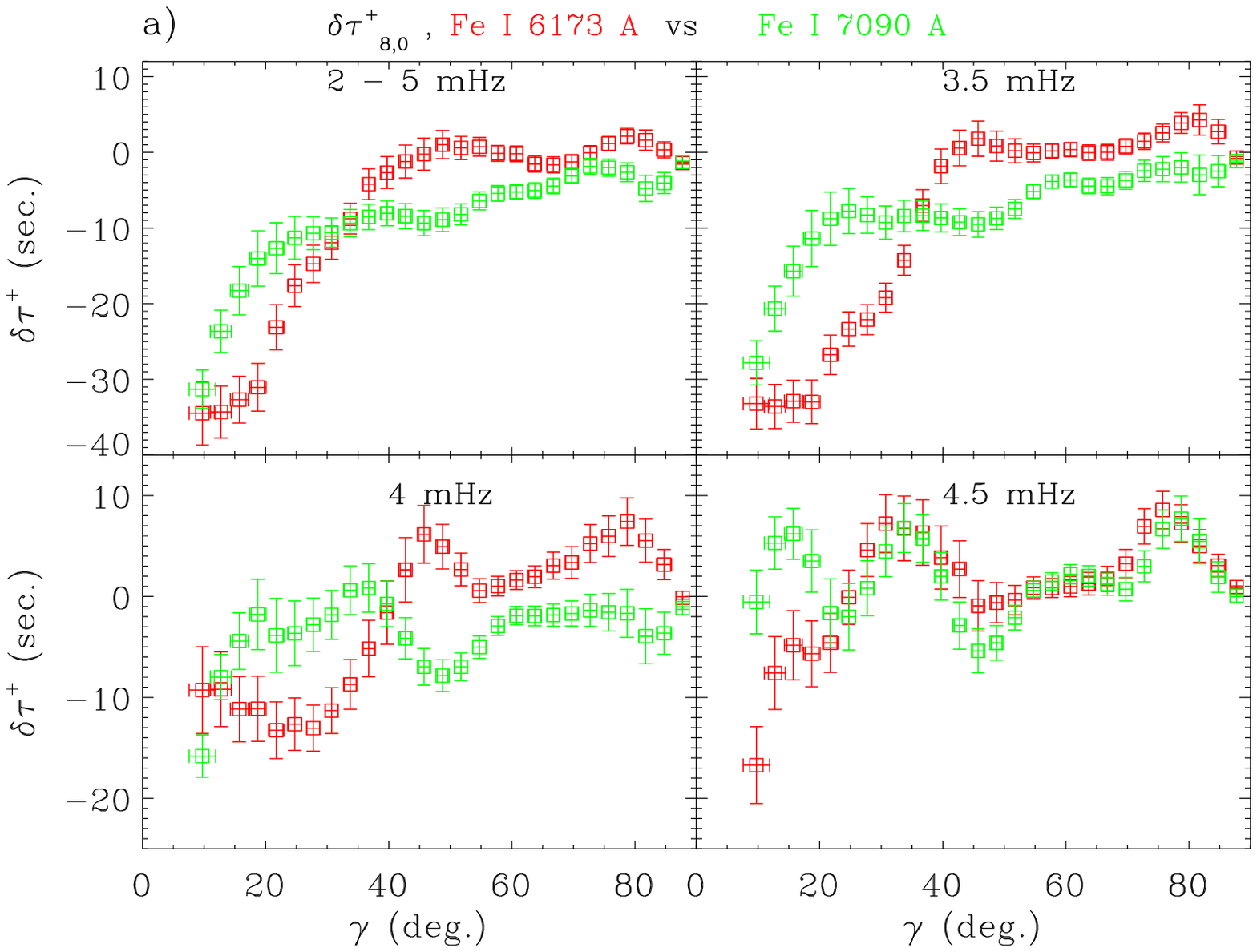}\qquad
            \includegraphics[width=7cm]{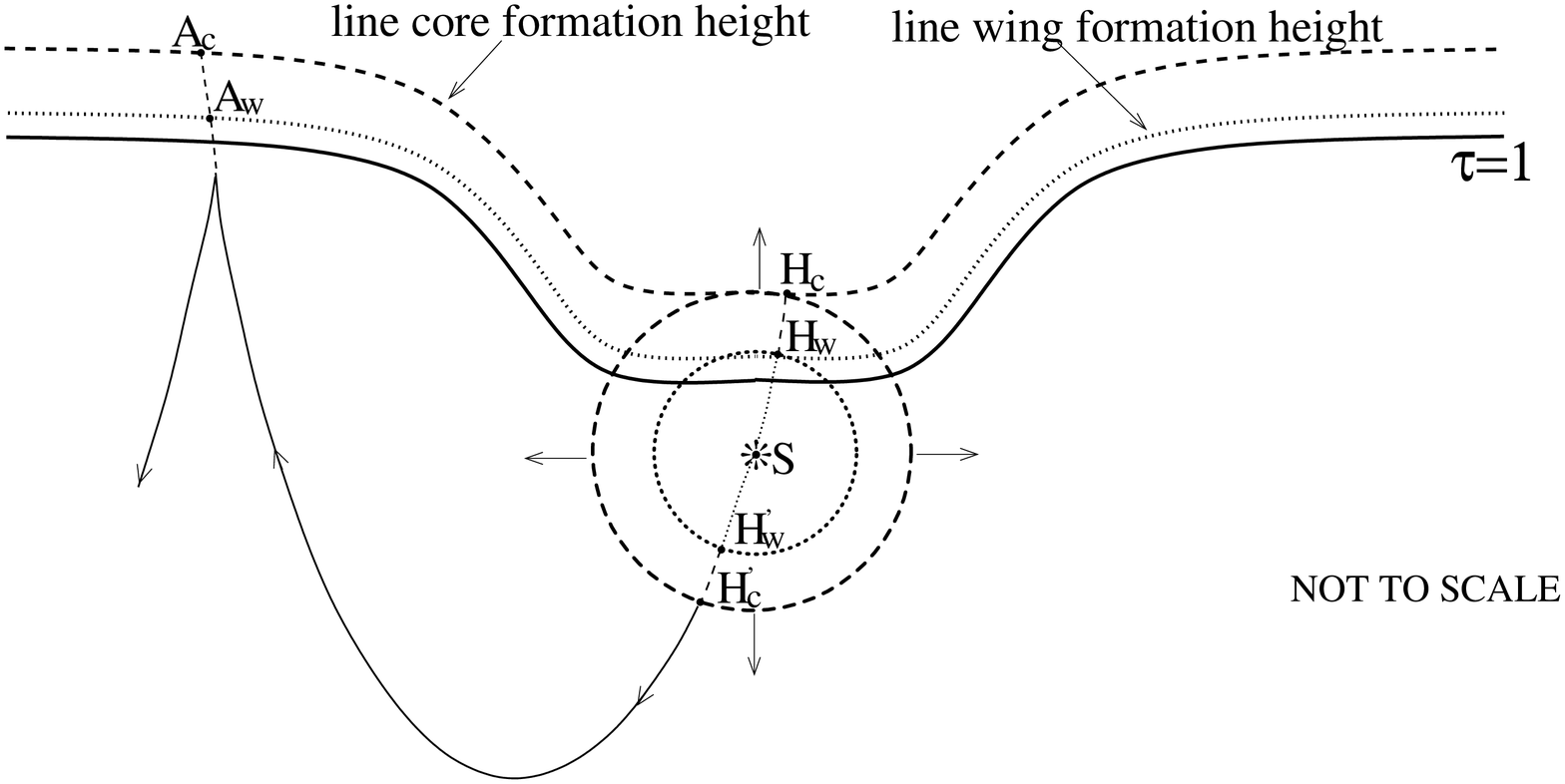}}
\caption{Changes in outgoing wave travel times, $\delta\tau^{+}_{8,0}$, due to wave propagation between the
formation heights of the wings (20 km) and the core (270 km) of Fe {\sc i} 6173 \AA~ (in red) and 7090 \AA~ (in green)
lines as a function of $\gamma$ of B. ({\em Panel b}), a cartoon depicting wavefronts from acoustic sources beneath the
umbra, wave paths and line formation heights (see text for details).}
\label{fig:3}
\end{figure}
Outgoing waves at a given measurement location, in general, would consist of those locally generated
and those generated elsewhere undergoing reflection at the photosphere directly below it. These latter component
would be seen in neither $\delta\phi_{i,0}$ nor $\delta\tau^{\pm}_{i,0}$, as they are evanescent at the observing height.
For locally generated waves, circular wavefronts from a source, while their upward propagating parts see
themselves up through the magnetic field, would cause outgoing wave correlations yielding distinct
signatures in $\delta\tau^{+}_{i,0}$ (see Figure 2(b)). Results in Figure 2(a), for $\delta\tau^{+}_{8,0}$ from
both the magnetic and non-magnetic lines, do indeed provide such a
diagnostic: outgoing waves starting at higher height (line core) within the sunspot atmosphere
and reaching the quiet-Sun at the chosen $\Delta$ have shorter travel times than those starting at a lower height
(line wings) and reaching
the same quiet-Sun location; since this is simply not possible, the only explanation for this observation
is the one contained in our previous sentence and illustrated in Figure 2(b),
viz., outgoing wave time-distance correlations are predominantly
due to waves directly from sources just beneath the sunspot photosphere when oscillations observed within it are used.

\section{Discussions and conclusion}

Almost all time-distance helioseismic analyses proceed under the working assumption that wave signals
at observation heights are evanescent and hence oppositely directed wave paths
involving photospheric reflections at two separated points are of identical path length. This assumption is basic to the
inferences on flows and wave speed from travel time differences and mean, respectively.
In an early theoretical study, accompanied by attempts to model the helioseismic observations
of \citet{braun97}, \citet{1998ApJ...492..379B} showed the influences of both the $p$-mode forcing of, and spontaneous emissions by,
sunspots on acoustic wave travel times.
Our analyses here have yielded transparent observational proofs for both effects, for the first time,
with important new perspectives: (1) the process of transformation of incident acoustic waves into propagating
(magneto)-acoustic waves up through the magnetic field happen in a coherent manner allowing a smooth evolution
of time-distance correlations and, in agreement with several recent theoretical and numerical studies 
\citep{cally05,2005SoPh..227....1C,schunkeretal06}, this process depends on the inclination angle 
($\gamma$) of magnetic field to the vertical,
and (2) outgoing waves from acoustic sources located just beneath the
sunspot photosphere add important additional contributions for both mean travel times and differences.
Our results have also shown observational prospects for
consistently accounting for the above effects in sunspot seismology, viz. the indispensability of imaging spectroscopy to
extract wave fields so as to be able to correctly account for the wave evolution within the directly
observable layers of sunspot atmosphere. 
%Current limitations in making such observations over large enough FOV
%do not allow us to perform seismic inversions reliably. However, the analysis methods followed here
%point ways to a consistent and much improved observational determinations of structure and flows beneath sunspots
%once our instrumental capabilities improve. These observational avenues also promise a close scrutiny of
%various theoretical ideas and models of acoustic wave - magnetic field interactions and those of the associated
%MHD waves and their propagation characteristics.

%\bibitem[Rajaguru, Basu \& Antia(2001)]{rajetal01}Rajaguru S. P., Basu, S., Antia, H.M. 2001, ApJ, 563, 410 

%\bibitem[Rajaguru et al.(2007)]{rajetal07}Rajaguru S. P., Sankarasubramanian K., Wachter R., Scherrer P. H., 2007, ApJ, 654, L175


\begin{thebibliography}{24}
\expandafter\ifx\csname natexlab\endcsname\relax\def\natexlab#1{#1}\fi

\bibitem[{{Basu} {et~al.}(2004){Basu}, {Antia}, \&
  {Bogart}}]{2004ApJ...610.1157B}
{Basu}, S., {Antia}, H.~M., {Bogart}, R.~S. 2004, {\it Astrophys. J.}, 610, 1157

\bibitem[{{Bogdan} {et~al.}(1998){Bogdan}, {Braun}, {Lites}, \&
  {Thomas}}]{1998ApJ...492..379B}
{Bogdan}, T.~J., {Braun}, D.~C., {Lites}, B.~W., {Thomas}, J.~H. 1998, {\it Astrophys. J.},
  492, 379
\bibitem[Braun(1997)]{braun97} Braun, D.C. 1997, {\it Astrophys. J.}, 487, 447

\bibitem[{{Braun} \& {Birch}(2006)}]{2006ApJ...647L.187B}
{Braun}, D.~C. {Birch}, A.~C. 2006, {\it Astrophys. J. Letters}, 647, L187

\bibitem[{{Braun} {et~al.}(1987){Braun}, {Duvall}, \&
  {Labonte}}]{1987ApJ...319L..27B}
{Braun}, D.~C., {Duvall}, Jr., T.~L., {Labonte}, B.~J. 1987, {\it Astrophys. J. Letters}, 319, L27

\bibitem[{{Braun} \& {Lindsey}(2000)}]{2000SoPh..192..307B}
{Braun}, D.~C. {Lindsey}, C. 2000, {\it Sol. Phys.}, 192, 307

\bibitem[{{Cally} {et~al.}(2003){Cally}, {Crouch}, \&
  {Braun}}]{2003MNRAS.346..381C}
{Cally}, P.~S., {Crouch}, A.~D., {Braun}, D.~C. 2003, {\it MNRAS}, 346, 381

\bibitem[Cally(2005)]{cally05} Cally, P.S. 2005, {\it MNRAS}, 358, 353

\bibitem[{{Crouch} \& {Cally}(2005)}]{2005SoPh..227....1C}
{Crouch}, A.~D. {Cally}, P.~S. 2005, {\it Sol. Phys.}, 227, 1

\bibitem[del Toro Iniesta (2003)]{dtiniesta03} del Toro Iniesta, J. C. 2003, Introduction to
Spectropolarimetry, Cambridge Univ. Press, p.149-164, Cambridge

\bibitem[{{Duvall} {et~al.}(1993){Duvall}, {Jefferies}, {Harvey}, \&
  {Pomerantz}}]{1993Natur.362..430D}
{Duvall}, Jr., T.~L., {Jefferies}, S.~M., {Harvey}, J.~W., {Pomerantz}, M.~A.
  1993, {\it Nature}, 362, 430

\bibitem[{{Duvall} {et~al.}(1996){Duvall}, {D'Silva}, {Jefferies}, {Harvey}, \&
  {Schou}}]{1996Natur.379..235D}
{Duvall}, T.~L.~J., {D'Silva}, S., {Jefferies}, S.~M., {Harvey}, J.~W.,
  {Schou}, J. 1996, {\it Nature}, 379, 235

\bibitem[{{Gizon} {et~al.}(2000){Gizon}, {Duvall}, \&
  {Larsen}}]{2000JApA...21..339G}
{Gizon}, L., {Duvall}, Jr., T.~L., {Larsen}, R.~M. 2000, Journal of
  Astrophysics and Astronomy, 21, 339

\bibitem[Gizon et al.(2010)]{gizonetal10} Gizon, L., Birch, A. C. \& Spruit, H. C. 2010, {\it Annual Rev. of Astronomy and 
Astrophysics}, vol. 48, p.289-338 

\bibitem[{{Hill}(1988)}]{1988ApJ...333..996H}
{Hill}, F. 1988, {\it Astrophys. J.}, 333, 996

\bibitem[{{Kosovichev} {et~al.}(2000){Kosovichev}, {Duvall}, \&
  {Scherrer}}]{2000SoPh..192..159K}
{Kosovichev}, A.~G., {Duvall}, T.~L.~.~J., {Scherrer}, P.~H. 2000, {\it Sol. Phys.},
  192, 159

\bibitem[{{Lindsey} \& {Braun}(1990)}]{1990SoPh..126..101L}
{Lindsey}, C. {Braun}, D.~C. 1990, {\it Sol. Phys.}, 126, 101

\bibitem[{{Lindsey} \& {Braun}(2005{\natexlab{a}})}]{2005ApJ...620.1107L}
{Lindsey}, C. {Braun}, D.~C. 2005{\natexlab{a}}, {\it Astrophys. J.}, 620, 1107

\bibitem[{{Lindsey} \& {Braun}(2005{\natexlab{b}})}]{2005ApJ...620.1118L}
{Lindsey}, C. {Braun}, D.~C. 2005{\natexlab{b}}, {\it Astrophys. J.}, 620, 1118

\bibitem[Moradi et al.(2010)]{2010SoPh..267....1M} Moradi, H., et al. 2010, {\it Sol. Phys.}, 267, 1

\bibitem[{{Parker}(1979)}]{1979ApJ...230..905P}
{Parker}, E.~N. 1979, {\it Astrophys. J.}, 230, 905

\bibitem[{{Rajaguru} {et~al.}(2001){Rajaguru}, {Basu}, \&
  {Antia}}]{2001ApJ...563..410R}
{Rajaguru}, S.~P., {Basu}, S., {Antia}, H.~M. 2001, {\it Astrophys. J.}, 563, 410

\bibitem[Rajaguru et al.(2004)]{rajaguruetal04} Rajaguru, S.P., Hughes, S.J. \& Thompson, M.J. 2004, {\it Sol. Phys.}, 220, 381

%\bibitem[Rajaguru et al.(2006)]{rajaguruetal06} Rajaguru, S.P., Birch, A.C., Duvall, T.L., Jr., Thompson, M.J.,
%\& Zhao, J. 2006, {\it Astrophys. J.}, 646, 543

\bibitem[Rajaguru et al.(2007)]{rajaguruetal07} Rajaguru, S.P., Sankarasubramanian, K., Wachter, R. \& Scherrer, P.H.
2007, {\it Astrophys. J.}, 654, L175

\bibitem[Rajaguru et al.(2010)]{rajaguruetal10} Rajaguru, S.P., Wachter, R., Sankarasubramanian, K. \& Couvidat, S.
2010, {\it Astrophys. J.}, 721, L86 
%\bibitem[Rimmele(1995)]{rimmele95} Rimmele, T. 1995, \aap, 298, 260
\bibitem[Sankarasubramanian \& Rimmele(2002)]{sankar-rimmele02} Sankarasubramanian, K. \& Rimmele, T. 2002,
{\it Astrophys. J.}, 576, 1048

\bibitem[{{Schunker} {et~al.}(2005){Schunker}, {Braun}, {Cally}, \&
  {Lindsey}}]{2005ApJ...621L.149S}
{Schunker}, H., {Braun}, D.~C., {Cally}, P.~S., {Lindsey}, C. 2005, {\it Astrophys. J. Letters}, 621,
  L149

\bibitem[Schunker \& Cally(2006)]{schunkeretal06} Schunker, H., \& Cally, P.S. 2006, {\it MNRAS}, 372, 551

\bibitem[{{Schunker} {et~al.}(2006){Schunker}, {Braun}, {Cally}, \&
  {Lindsey}}]{2006ASPC..354..244S}
{Schunker}, H., {Braun}, D.~C., {Cally}, P.~S., {Lindsey}, C. 2006, in Solar
  MHD Theory and Observations: A High Spatial Resolution Perspective, eds.
  J.~{Leibacher}, R.~F. {Stein}, \& H.~{Uitenbroek}, Astronomical Society of
  the Pacific Conference Series, 354, 244

\bibitem[Svanda et al.(2011)]{2011A&A...530A.148S} Svanda, M., Gizon, L., Hanasoge, S.M.,  Ustyugov, S.D 2011, 
{\it Astron. \& Astrophys.}, 530, 148
 
%\bibitem[{{Spruit}(1996)}]{1996BASI...24..211S}
%{Spruit}, H.~C. 1996, Bulletin of the Astronomical Society of India, 24, 211

%\bibitem[{{Spruit} \& {Bogdan}(1992)}]{1992ApJ...391L.109S}
%{Spruit}, H.~C. {Bogdan}, T.~J. 1992, {\it Astrophys. J. Letters}, 391, L109

\bibitem[{{Thomas} {et~al.}(1982){Thomas}, {Cram}, \&
  {Nye}}]{1982Natur.297..485T}
{Thomas}, J.~H., {Cram}, L.~E., {Nye}, A.~H. 1982, {\it Nature}, 297, 485

%\bibitem[{{Woodard}(1997)}]{1997ApJ...485..890W}
%{Woodard}, M.~F. 1997, {\it Astrophys. J.}, 485, 890

\bibitem[{{Zhao} {et~al.}(2001){Zhao}, {Kosovichev}, \&
  {Duvall}}]{2001ApJ...557..384Z}
{Zhao}, J., {Kosovichev}, A.~G., {Duvall}, Jr., T.~L. 2001, {\it Astrophys. J.}, 557, 384

\end{thebibliography}
\end{document}